\documentclass[twocolumn,showpacs,preprintnumbers,amsmath,amssymb,APSl,prd,nofootinbib,superscriptaddress]{revtex4-1}

\usepackage{dcolumn}
\usepackage{bm}
\usepackage{ifpdf}
\usepackage{hyperref}
\usepackage{bm}
\usepackage{xcolor,color,graphicx,graphics}
\usepackage[spanish,english]{babel}
\usepackage[latin1]{inputenc}
\usepackage[OT1]{fontenc}
\usepackage{latexsym,amssymb,amsmath,amsfonts}
\usepackage{makeidx}
\usepackage{epsfig,subfigure}
\usepackage{natbib}
\usepackage{epstopdf}
\usepackage{mathrsfs}
\usepackage{hyperref}

\newcommand{\be}{\begin{equation}}
\newcommand{\en}{\end{equation}}
\newcommand{\bea}{\begin{eqnarray}}
\newcommand{\ena}{\end{eqnarray}}
\newcommand{\bes}{\begin{subequations}}
\newcommand{\ees}{\end{subequations}}

\begin{document}

\title{Metric-affine $f(R,T)$ theories of gravity and their applications}

\author{E. Barrientos} \email{ebarrientos@astro.unam.mx}
\affiliation{Instituto de Astronom\'ia, Universidad Nacional Aut\'onoma de M\'exico, \\
AP 70-264, Ciudad de M\'exico, 04510, M\'exico}

\author{Francisco S. N. Lobo} \email{fslobo@fc.ul.pt}
\affiliation{Instituto de Astrof\'{\i}sica e Ci\^{e}ncias do Espa\c{c}o, Faculdade de
Ci\^encias da Universidade de Lisboa, Edif\'{\i}cio C8, Campo Grande,
P-1749-016 Lisbon, Portugal}

\author{S. Mendoza} \email{sergio@astro.unam.mx}
\affiliation{Instituto de Astronom\'ia, Universidad Nacional Aut\'onoma de M\'exico, \\
AP 70-264, Ciudad de M\'exico, 04510, M\'exico}

\author{Gonzalo J. Olmo} \email{gonzalo.olmo@uv.es}
\affiliation{Departamento de F\'{i}sica Te\'{o}rica and IFIC, Centro Mixto Universidad de Valencia - CSIC.
Universidad de Valencia, Burjassot-46100, Valencia, Spain}
\affiliation{Departamento de F\'isica, Universidade Federal da
Para\'\i ba, 58051-900 Jo\~ao Pessoa, Para\'\i ba, Brazil}

\author{D. Rubiera-Garcia} \email{drgarcia@fc.ul.pt}
\affiliation{Instituto de Astrof\'{\i}sica e Ci\^{e}ncias do Espa\c{c}o, Faculdade de
Ci\^encias da Universidade de Lisboa, Edif\'{\i}cio C8, Campo Grande,
P-1749-016 Lisbon, Portugal}

\date{\today}
\begin{abstract}
We study $f(R,T)$ theories of gravity, where $T$ is the trace of the energy-momentum tensor $T_{\mu\nu}$, with independent metric and affine connection (metric-affine theories). We find that the resulting field equations share a close resemblance with their metric-affine $f(R)$ relatives once an effective energy-momentum tensor is introduced. As a result, the metric field equations are second-order and no new propagating degrees of freedom arise as compared to GR, which contrasts with the metric formulation of these theories, where a dynamical scalar degree of freedom is present. Analogously to its metric counterpart, the field equations impose the nonconservation of the energy-momentum tensor, which implies nongeodesic motion and consequently leads to the appearance of an extra force. The weak field limit leads to a modified Poisson equation formally identical to that found in Eddington-inspired Born-Infeld gravity. Furthermore, the coupling of these gravity theories to perfect fluids, electromagnetic, and scalar fields, and their potential applications are discussed.
\end{abstract}

\pacs{04.20.Dw, 04.40.Nr, 04.50.Kd, 04.70.Bw}

\maketitle

\section{Introduction}

Modified theories of gravity are a mainstream topic in modern cosmology, essentially due to the discovery of the late-time cosmic accelerated expansion \cite{Perlmutter:1998np,Riess:1998cb}. These theories assume that Einstein's General Relativity (GR) breaks down at large scales and that an extension of the Einstein-Hilbert action describing the gravitational field is necessary, offering an alternative paradigm fundamentally distinct from dark energy models of cosmic acceleration \cite{Copeland:2006wr,Nojiri:2017ncd}. Further physical motivations for these theories include a more realistic representation of quantum and gravitational fields at high-energy densities near curvature singularities, and the possibility to create some effective first order approximation of quantum gravity \cite{PT,BOS}. The simplest such extension of GR is perhaps to consider a Lagrangian density given by a certain function $f(R)$, where $R$ is the scalar curvature, whose phenomenology has been largely explored in the literature \cite{modgrav,modgrav2,modgrav3}.

An interesting generalization of $f(R)$ gravity involves the inclusion of a nonminimal coupling between the scalar curvature and matter \cite{HGoenner,Bertolami:2007gv,Nojiri:2004bi,Allemandi:2005qs}. One of the original motivations to implement this coupling was to establish a link with MOND and the flat galactic rotation curves. It was further shown that this curvature-matter coupling induces a non-vanishing covariant derivative of the energy-momentum tensor, which implies nongeodesic motion and consequently leads to the appearance of an extra force \cite{Bertolami:2007gv}. Thus, these models allow for an explicit violation of the equivalence principle (EP), which is tightly constrained by solar system experimental tests \cite{Bertolami:2006js}, by imposing a matter-dependent deviation from geodesic motion. Low-energy features of specific compactified versions of higher-dimensional theories also imply the EP violation \cite{Overduin:2000gr}. However, it has been argued that the EP is not one of the ``universal'' principles of physics \cite{Damour:2001fn}, but rather it is a heuristic hypothesis introduced by Einstein, and used to construct his theory of GR. Further tests of the EP are relevant for new physics and strongly constrain the parameters of the theory \cite{Damour:1996xt, Damour:2010rp}. However, it is important to note that the violation of the EP does not in principle rule out the specific theory.

The linear nonminimal curvature-matter coupling \cite{Bertolami:2007gv} was further generalized by considering a maximal extension of the Einstein-Hilbert action, namely, $f(R,\mathcal{L}_m)$ gravity  \cite{Harko:2010mv}, where $\mathcal{L}_m$ is the matter Lagrangian. A related theory is $f(R,T)$ gravity, where the gravitational Lagrangian is given by an arbitrary function of the Ricci scalar and the trace $T$ of the energy-momentum tensor $T_{\mu\nu}$ \cite{Harko:2011kv}. All of these theories induce the presence of an extra force and consequently nongeodesic motion. An interesting cosmological motivation for $f(R,T)$ gravity is that it may be considered a relativistically covariant model of interacting dark energy \cite{Harko:2011kv}. Note that the dependence from $T$ may be induced by exotic imperfect fluids or quantum effects (conformal anomaly). A physical interpretation consists on the possibility that the curvature-matter coupling is related to the thermodynamics of open systems, and is responsible for matter creation irreversible processes that may take place at a cosmological scale \cite{Harko:2014pqa,Harko:2015pma}. Fundamental applications of the curvature-matter couplings in the study of quantum gravitational theories with first order quantum corrections induced by a stochastically fluctuating metric have also been analysed \cite{Liu:2016qfx}.
It is interesting to note that in recent work \cite{Avelino:2018qgt,Avelino:2018rsb}, it was argued that the on-shell Lagrangian of a perfect fluid depends on microscopic properties of the fluid, and consequently it was shown that if the fluid is constituted by localized concentrations of energy with fixed rest mass and structure (solitons) then the average on-shell Lagrangian of a perfect fluid is given by $\mathcal{L}_m=T$. Thus, this seems to indicate that, in this context, $f(R,\mathcal{L}_m)$ theories may be regarded as a subclass of $f(R,T)$ gravity. Further arguments in favor of these theories are found on the fact that the relativistic behavior of a Tully-Fisher law observed in the rotation of galaxies can be modelled with a $f(R,T)$ or $f(R,\mathcal{L}_m)$ description, as shown in \cite{mendoza15}, which is coherent with lensing observations of individual, groups and clusters of galaxies. The literature of $f(R,T)$ gravity is extremely vast and we refer the reader to the review \cite{Harko:2014gwa} for further motivations and applications.

The current approach to $f(R,T)$ theories is framed within the so-called metric formulation, where the affine structure of the spacetime geometry is dictated by the metric tensor\footnote{It has been argued in \cite{Alvarenga:2013syu} that in this approach models of the form $f(R,T)=f_1(R)+f_2(T)$ yield a scale-dependent behavior of scalar cosmological perturbations that is disfavored by observational data, severely limiting the viability of such models.}. Other approaches, however, are possible. In fact, if one allows the connection to vary independently of the metric tensor, the so-called metric-affine or Palatini approach, the resulting field equations typically lead to different dynamics, offering alternative avenues to explore new gravitational physics. The curvature-matter coupling in metric-affine approach has been scarcely considered in the literature \cite{Harko:2010hw}, with the main highlight being that the independent connection can be expressed as the Levi-Civita connection of an auxiliary (matter Lagrangian-dependent) metric, which is related with the physical metric by means of a conformal transformation. Analogously to the metric case \cite{Bertolami:2007gv}, the field equations impose the nonconservation of the energy-momentum tensor. In this framework, the FLRW equations for brane-world cosmology and loop quantum cosmology can be derived out of a quadratic $f(R)$ theory plus a nonminimal linear coupling between matter and curvature \cite{Olmo:2014sra}. Let us also point out that generalized descriptions of galaxies rotation curves have been previously implemented in the literature using a metric-affine formalism with torsion included in the description of the gravitational action \cite{barrientos16,barrientos17}.

The main aim of this work is to address in detail $f(R,T)$ theories in this, so far quite unexplored, alternative metric-affine view. We will show that the resulting theories are radically different in some aspects from their metric counterparts, though they share many resemblances with their $f(R)$ relatives. In fact, the study of modified theories of gravity in metric-affine scenarios involving torsion and nonmetricity has received a continuous interest in the last two decades, with several review articles focused on those topics \cite{Hehl:1994ue,OlmoIJMPD,Shapiro:2001rz}. This work will pave the path for future studies of $f(R,T)$ theories in geometric scenarios where torsion and nonmetricity are not \textit{a priori} constrained to vanish. We note in this regard that whether the spacetime structure is Riemannian or otherwise is a foundational question of gravitational physics that must be answered empirically, not decided by convention or on practical terms.

This paper is organized as follows: In Sec. \ref{sec:eom}, we present the formalism of $f(R,T)$ gravity in the metric-affine approach, focussing on the role of the curvature-matter coupling in the equations of motion, the conservation equation, and the geodesic motion and presence of a fifth force. In Sec. \ref{sec:wf}, we trace out the weak field limit and show that the modified Poisson equation is formally identical to that found in Eddington-inspired Born-Infeld gravity. In Sec. \ref{sec:mod}, we present several specific applications, such as the stellar structure equations, and in the presence of electromagnetic fields and scalar fields. Finally, in Sec. \ref{sec:con}, we summarize our results and depict some future applications.

\section{Theory, formulation, and equations of motion} \label{sec:eom}

To introduce the action of $f(R,T)$ gravity in the metric-affine approach one needs to bear in mind that only the affine connection $\Gamma_{\mu\nu}^{\lambda}$ is needed to define the Ricci tensor, which follows from the Riemann tensor
\begin{equation}\label{eq:Riemann}
{R^\alpha}_{\beta\mu\nu}=\partial_{\mu}
\Gamma^{\alpha}_{\nu\beta}-\partial_{\nu}
\Gamma^{\alpha}_{\mu\beta}+\Gamma^{\alpha}_{\mu\lambda}\Gamma^{\lambda}_{\nu\beta}-\Gamma^{\alpha}_{\nu\lambda}\Gamma^{\lambda}_{\mu\beta} \ ,
\end{equation}
as $R_{\mu\nu}(\Gamma) \equiv {R^\alpha}_{\mu\alpha\nu}(\Gamma)$ (no indices lowered/raised with the metric). Subsequent contraction with the metric $g_{\mu\nu}$ allows to define the curvature scalar as $R \equiv g^{\mu\nu}R_{\mu\nu}(\Gamma)$. This guarantees that only the symmetric part of the Ricci tensor enters into the action, which significantly simplifies the role of torsion, making it irrelevant if fermions are not considered \cite{Afonso:2017bxr}. Throughout this work, we assumme the $(-,+,+,+)$ signature. With these elements the action considered in this work takes the form
\begin{equation} \label{eq:action}
\mathcal{S}=\frac{1}{2\kappa^2} \int d^4x \sqrt{-g} f(R,T) + \int d^4x \sqrt{-g} \mathcal{L}_m(g_{\mu\nu},\psi_m) \ ,
\end{equation}
with the following definitions and conventions: $\kappa^2$ is some constant with suitable dimensions (in GR, $\kappa^2=8\pi G$), $g$ is the determinant of the spacetime metric $g_{\mu\nu}$, the factor $f(R,T)$ is an arbitrary function of the curvature scalar $R$ and the trace of the energy-momentum tensor, $T \equiv g^{\mu\nu} T_{\mu\nu}$, which is defined as
\begin{equation}\label{eq:emtensor}
T_{\mu\nu}=-\frac{2}{\sqrt{-g}} \frac{\delta (\sqrt{-g} \mathcal{L}_m) }{\delta g^{\mu\nu}} \ .
\end{equation}
Finally, the standard matter Lagrangian density $\mathcal{L}_m$ depends on the matter fields $\psi_m$ and the metric $g_{\mu\nu}$, but not on the independent connection $\Gamma_{\mu\nu}^{\lambda}$.

The variation of the action (\ref{eq:action}) can be conveniently expressed as
\begin{eqnarray} \label{eq:variation}
\delta \mathcal{S}&=&\int \frac{d^4x \sqrt{-g}}{2\kappa^2}  \Big[ f_R R_{\mu\nu} - \frac{1}{2}g_{\mu\nu}f + f_T \frac{\delta T}{\delta g^{\mu\nu}} - \kappa^2 T_{\mu\nu} \Big]\delta g^{\mu\nu} \nonumber \\
&+&\frac{1}{2\kappa^2} \int d^4x \sqrt{-g}  f_R g^{\mu\nu} \delta R_{\mu\nu}(\Gamma)\,,
\end{eqnarray}
where we have defined $f_R \equiv df/dR$ and $f_T \equiv df/dT$ and split the variation into two lines to highlight the variations with respect to the metric and with respect to the affine connection, respectively. Now using the fact that the variation of $T$ with respect to $g_{\mu\nu}$ can be written as
\begin{equation}
\frac{\delta T}{\delta g^{\mu\nu}}=T_{\mu\nu}+\Theta_{\mu\nu} \,,
\end{equation}
where
\begin{equation} \label{eq:theta}
\Theta_{\mu\nu} \equiv g^{\lambda \rho} \frac{\delta T_{\lambda \rho}}{\delta g^{\mu\nu}} \ ,
\end{equation}
then the variation of Eq. (\ref{eq:variation}) with respect to $g_{\mu\nu}$ can be expressed as
\begin{equation}\label{eq:metric}
f_R R_{\mu\nu} - \frac{1}{2} g_{\mu\nu}f=\kappa^2 \tau_{\mu\nu} \,,
\end{equation}
where we have introduced the \emph{effective} energy-momentum tensor
\begin{equation} \label{eq:tau}
\tau_{\mu\nu}=T_{\mu\nu}\left(1-\frac{f_T}{\kappa^2} \right) - \frac{f_T}{\kappa^2}  \Theta_{\mu\nu} \,,
\end{equation}
which plays a key role in the dynamics of these models, as shall be clear later. On the other hand, from the variation of the Ricci tensor in (\ref{eq:variation}), after integration by parts and a bit of algebra one finds\footnote{For a detailed derivation of these equations including torsion, see \cite{Afonso:2017bxr}. }
\begin{equation} \label{eq:connection}
\nabla_{\lambda}^{\Gamma}(\sqrt{-g} f_R g^{\mu\nu})=0 \,.
\end{equation}
The two sets of Eqs. (\ref{eq:metric}) and (\ref{eq:connection}) can be written in a more suitable form by noting that the contraction of (\ref{eq:metric}) with the metric $g^{\mu\nu}$ yields the result
\begin{equation} \label{eq:traceEq}
R f_R  - 2f=\kappa^2 \tau \,,
\end{equation}
where $\tau \equiv g^{\mu\nu} \tau_{\mu\nu}$. Note that (\ref{eq:traceEq}) is an algebraic equation rather than a differential one and implies that, like in the metric-affine $f(R)$ case, the curvature scalar is a function of the matter sources only. This allows to introduce a new rank-two tensor $h_{\mu\nu}$ such that the connection equations (\ref{eq:connection}) can be expressed as $\nabla_{\lambda}^{\Gamma}(\sqrt{-h} h^{\mu\nu})=0$, which implies the conformal relation
\begin{equation} \label{eq:conf}
h_{\mu\nu}=f_R g_{\mu\nu} \,,
\end{equation}
between these two metrics. This way, the affine connection $\Gamma_{\mu\nu}^{\lambda}$ is given by the Christoffel symbols of the metric $h_{\mu\nu}$, i.e.,
\begin{equation}
\Gamma^\lambda_{\mu\nu}= \frac{h^{\lambda\alpha}}{2}\left(\partial_\mu h_{\alpha\nu}+\partial_\nu h_{\alpha\mu}-\partial_\alpha h_{\mu\nu}\right) \,.
\end{equation}

Now, contracting Eqs. (\ref{eq:metric}) with $h^{\alpha\mu}$, using the conformal relation (\ref{eq:conf}), and rearranging terms one arrives at
\begin{equation} \label{eq:Riccieqs}
{R^\mu}_{\nu}(h)=\frac{\kappa^2}{f_R^2} \left({\tau^\mu}_{\nu} + \frac{f(R,T)}{2\kappa^2} {\delta^\mu}_{\nu} \right) \,,
\end{equation}
where ${R^\mu}_{\nu}(h) \equiv h^{\mu\alpha}R_{\alpha \nu}$. Written in this form, Eqs. (\ref{eq:Riccieqs}) become (for any $f(R,T)$ function) a system of second-order differential Einstein-like field equations for the metric $h_{\mu\nu}$, with all the terms on the right-hand side being functions of the matter sources, and representing a natural generalization of metric-affine $f(R)$ theories with the $f_T$-corrections encoded in the effective energy-momentum tensor ${\tau^\mu}_{\nu}$ of Eq. (\ref{eq:tau}). After solving these equations for $h_{\mu\nu}$ one just needs to use the conformal relation (\ref{eq:conf}) to find the spacetime metric $g_{\mu\nu}$. A corollary of these features is that, in vacuum, ${T_\mu}^{\nu}=0$, all the terms on the right-hand side vanish, one finds that $h_{\mu\nu}=g_{\mu\nu}$ (modulo a trivial rescaling), and the same vacuum solutions of GR (with possibly a cosmological constant term) are recovered. This implies that the propagating degrees of freedom  present in these theories are the same as those in GR.

\subsection{The role of the curvature-matter coupling}

To fully specify these theories of gravity one needs not only the particular dependence on the scalar curvature but also the matter Lagrangian density $\mathcal{L}_m$. Once the latter is given, one can compute explicitly the object $\Theta_{\mu\nu}$ in Eq.(\ref{eq:theta}) as \cite{Harko:2011kv}
\begin{equation} \label{eq:Thetaexp}
\Theta_{\mu\nu}=-2T_{\mu\nu} +g_{\mu\nu} \mathcal{L}_m -2g^{\alpha\beta} \frac{\partial^2 \mathcal{L}_m}{\partial g^{\alpha \beta} \partial g^{\mu\nu}} \,.
\end{equation}
This expression allows to rewrite ${\tau^\mu}_{\nu}$ in Eq. (\ref{eq:tau}) into the more suggestive form
\begin{equation}\label{eq:tau2}
{\tau^\mu}_{\nu}=T_{(I)\nu}^{\mu}+\frac{f_T}{\kappa^2}\left[T_{(I) \nu}^{\mu}-T_{(II)\nu}^{\mu}\right] \,,
\end{equation}
where for convenience we have introduced the tensors
\begin{eqnarray}
T_{(I)\nu}^{\mu}&=&-2g^{\mu\rho}\frac{\partial \mathcal{L}_m}{\partial g^{\rho \nu}}+\mathcal{L}_m \delta^{\mu}_{\nu} \,, \label{TmunuI} \\
T_{(II)\nu}^{\mu}&=&-2g^{\mu\rho}g^{\alpha\beta}\frac{\partial^2 \mathcal{L}_m}{\partial g^{\alpha\beta}\partial g^{\rho \nu}}+\mathcal{L}_m \delta^{\mu}_{\nu} \,.
\end{eqnarray}
The first one corresponds to the standard energy-momentum tensor defined in Eq. (\ref{eq:emtensor}), while the second one is a generalization involving second metric derivatives of the matter Lagrangian density. This structure suggests that it should be possible to consider more general theories containing additional couplings between gravity and the matter fields in this context. In particular, a family of $f(R,\tau)$ theories, with $\tau\equiv {\tau^\mu}_\mu$, would lead to an extension involving terms with three derivatives of $\mathcal{L}_m$ with respect to the metric, and so on.

A case of general interest for the matter fields is represented by a perfect fluid, whose energy-momentum tensor is  of the form
\begin{equation} \label{eq:pftumunu}
T_{\mu\nu}=(\rho + P) u_{\mu}u_{\nu}+P g_{\mu\nu} \,,
\end{equation}
where $u^{\mu}$ is the unit timelike vector, $u_{\mu}u^{\mu}=-1$, while $\rho$ and $P$ are the energy density and pressure of the fluid, respectively. For this matter source, we assume that $\mathcal{L}_m=P$ as the matter Lagrangian density \footnote{For an extended discussion on the well known problem of whether $\mathcal{L}_m=P$ or $\mathcal{L}_m=-\rho$ is the right Lagrangian of a perfect fluid, and its consequences for nonminimally coupled theories see e.g. \cite{Faraoni:2009rk,Bertolami:2008ab}.} which, from Eq. (\ref{eq:Thetaexp}), yields  \begin{equation}
\Theta_{\mu\nu}=-2 T_{\mu\nu}-Pg_{\mu\nu} \ .
\end{equation}
Inserting this result in Eq. (\ref{eq:Riccieqs}), one finds
\begin{equation} \label{eq:RiccieqsFluid}
{R^\mu}_{\nu}(h)=\frac{1}{f_R^2} \left[(\kappa^2+f_T){T^\mu}_{\nu} + \left(\frac{f}{2}+Pf_T\right) {\delta^\mu}_{\nu} \right] .
\end{equation}
From this expression, it is easy to verify that the limit $P\to 0$ recovers the same dynamics as metric-affine $f(R)$ theories but with a varying effective Newton's constant, namely, $\kappa^2_{eff}=\kappa^2+f_T$, with $f_T$ a function of $\rho$. If we further restrict to the case $f_T=$constant, then the correspondence is exact. This puts forward that the family of models $f(R,T)=f(R)+\epsilon T$ only departs from the $f(R)$ case in scenarios where the fluid pressure becomes relevant as compared to the term $f(R,T)/2$.

\subsection{Conservation equation}

Let us now work out the analogous of the conservation equation in these theories. First we rewrite the field equations (\ref{eq:Riccieqs}) as
\begin{equation}
{G^\mu}_{\nu}(h)=\frac{\kappa^2}{f_R^2}\left[{\tau^\mu}_{\nu}-\frac{\delta^{\mu}_{\nu}}{2} \left(\tau + \frac{f}{\kappa^2}\right)\right] \,.
\end{equation}
Taking a covariant derivative on both sides on this equation and using Bianchi's identities, $\nabla_{\mu}^{(h)} {G^\mu}_{\nu}(h) \equiv 0$ (the superindex $h$ indicates covariant derivatives defined with the independent connection $\Gamma_{\mu\nu}^{\lambda}$), one finds
\begin{eqnarray}
\nabla_{\mu}^{(h)}{\tau^\mu}_{\nu}&-&\frac{1}{2} \partial_{\nu}\left(\tau + \frac{f}{\kappa^2}\right) \nonumber \\
&-&2\partial_{\mu} \ln f_R \left[{\tau^\mu}_{\nu}-\frac{\delta^{\mu}_{\nu}}{2}\left( \tau + \frac{f}{\kappa^2} \right) \right]=0 \,.
\label{eq:coneq1}
\end{eqnarray}
On the other hand, the relation between covariant derivatives defined with the independent connection and those defined with the connection associated to the Christoffel symbols of the metric, $\nabla_{\mu}^{(g)}$, is obtained as
\begin{equation}
\nabla_{\mu}^{(h)}{\tau^\mu}_{\nu}=\nabla_{\mu}^{(g)}{\tau^\mu}_{\nu}+C_{\mu\lambda}^{\mu}{\tau^\lambda}_{\mu}-C_{\mu\nu}^{\lambda}{\tau^\mu}_{\lambda} \,,
\end{equation}
where
\begin{equation}
C_{\mu\nu}^{\alpha}=\frac{h^{\alpha\rho}}{2}\left[\nabla_{\mu}^{(g)}h_{\rho\nu}+\nabla_{\nu}^{(g)}h_{\rho\mu}-\nabla_{\rho}^{(g)}h_{\mu\nu}\right] \,.
\end{equation}
Now, using the conformal relation (\ref{eq:conf}) and after a bit of algebra upon the relation above one arrives at
\begin{equation}
\nabla_{\mu}^{(h)}{\tau^\mu}_{\nu}=\nabla_{\mu}^{(g)}{\tau^\mu}_{\nu}+2{\tau^\lambda}_{\nu}\partial_{\lambda}\ln f_R-\frac{\tau}{2}\partial_{\nu} \ln f_R \,.
\end{equation}
Plugging this result into the nonconservation equation (\ref{eq:coneq1}) yields
\begin{equation}
\nabla_{\mu}^{(g)}{\tau^\mu}_{\nu}+\left(\frac{\tau}{2}+\frac{f}{\kappa^2}\right) \frac{\partial_{\nu} f_R}{f_R}-\partial_{\nu}\left(\frac{\tau}{2}+\frac{f}{2\kappa^2}\right)=0 \,.
\end{equation}
Using now the trace equation (\ref{eq:traceEq}) to consider the combinations
\begin{eqnarray}
\frac{1}{2}\left(\tau + \frac{f}{\kappa^2}\right)&=&\frac{1}{2\kappa^2}(Rf_R-f) \,,
\\
\frac{\tau}{2} + \frac{f}{\kappa^2}&=&\frac{1}{2\kappa^2}Rf_R \,,
\end{eqnarray}
and after some manipulations we finally obtain the result
\begin{equation} \label{eq:contau}
\nabla_{\mu}^{(g)}{\tau^\mu}_{\nu}= -\frac{f_T}{2\kappa^2}\partial_\nu T \,,
\end{equation}
implying that  the effective energy-momentum tensor ${\tau^\mu}_{\nu}$ is conserved only when the term $f_T \partial_\nu T$ vanishes.  This has nontrivial consequences regarding several contexts, in particular, stellar structure, as shall be seen in Sec.~\ref{sec:mod} below.

\subsection{Geodesic equation and extra force}

In order to compute the geodesic equation obtained from the nonconservation
equation \eqref{eq:contau}, let us substitute the relation \eqref{eq:Thetaexp}
into the definition of $\tau_{\mu\nu}$ given by Eq. \eqref{eq:tau}, to
obtain
\begin{equation}\label{eq:taugeo}
\tau_{\mu\nu}=T_{\mu\nu}+2\frac{f_T}{\kappa^2}\left(g^{\alpha\beta}\frac{\partial^2 \mathcal{L}_m}{\partial g^{\alpha\beta}\partial g^{\mu\nu}}-\frac{\partial \mathcal{L}_m}{\partial g^{\mu\nu}}\right),
\end{equation}
where we have used the expression of the energy-momentum tensor given by Eq. (\ref{TmunuI}). Therefore,  Eq. (\ref{eq:contau}) implies that
\begin{eqnarray}\label{eq:divergencia}
\nabla_{\mu}^{(g)}{T^\mu}_{\nu}&=&\frac{2}{\kappa^2}\nabla_{\mu}^{(g)}\left[f_T g^{\mu\lambda} \left(\frac{\partial \mathcal{L}_m}{\partial g^{\lambda\nu}}-g^{\alpha\beta}\frac{\partial^2 \mathcal{L}_m}{\partial g^{\alpha\beta}\partial g^{\lambda\nu}}\right)\right] \nonumber \\
	&&-\frac{f_T}{2\kappa^2}\partial_\nu T.
\end{eqnarray}

On the other hand, since the matter current conservation relation $\nabla_{\mu}^{(g)}(\rho
u^\mu)=0$ implies that the quantity $u^\mu\rho\sqrt{-g}$ is conserved,
therefore the differential of this quantity is null. With this and using
the fact that \( 2 \delta u^\mu= u_\nu\delta g^{\mu\nu} \) and
\( 2 \delta \sqrt{-g}=\sqrt{-g}g_{\mu\nu}\delta g^{\mu\nu} \)
we obtain the following relation:
\begin{equation}\label{eq:rhog}
\delta \rho=\frac{1}{2}\rho(g_{\mu\nu} + u_\mu u_\nu)\delta g^{\mu\nu},
\end{equation}
which facilitates the computation of ${\partial \mathcal{L}_m}/{\partial
g^{\lambda\nu}}$ and ${\partial^2 \mathcal{L}_m}/{\partial g^{\alpha\beta}\partial g^{\lambda\nu}}$ on the right-hand side of (\ref{eq:divergencia}). With this last expression, the energy-momentum tensor (\ref{TmunuI}) is given by
\begin{equation}\label{Trho}
T_{\mu\nu}= -\rho u_\mu u_\nu\frac{d \mathcal{L}_m}{d\rho}+g_{\mu\nu}\left(\mathcal{L}_m-\rho\frac{d \mathcal{L}_m}{d\rho}\right).
\end{equation}
Using Eq. \eqref{eq:rhog} to express the derivatives of the matter Lagrangian with respect to the metric as derivatives with respect to $\rho$ in Eq. \eqref{eq:divergencia} yields
\begin{eqnarray}\label{eq:div1}
&&\nabla_{\mu}^{(g)}\left[-\rho u^\mu u_\nu\frac{d \mathcal{L}_m}{d\rho}+\delta^\mu_\nu\left(\mathcal{L}_m-\rho\frac{d \mathcal{L}_m}{d\rho}\right)\right]=
\nonumber \\
&&\nabla_{\mu}^{(g)}\left[\frac{f_T}{2\kappa^2}\left(\rho(u^\mu u_\nu+\delta^\mu_\nu)\left(\frac{d \mathcal{L}_m}{d\rho}-3\rho\frac{d^2 \mathcal{L}_m}{d\rho^2}\right)\right)\right] \nonumber \\
&&-\frac{f_T}{2\kappa^2}\partial_\nu T.
\end{eqnarray}

Now, by taking the divergences in the previous relation and recalling the well-known relation
\begin{equation}\label{cuadrivel}
u^\nu\nabla_{\nu}^{(g)}u^\mu=\frac{d^2x^\mu}{ds^2}+\Gamma^\mu\,_{\lambda\nu}\frac{dx^\lambda}{ds}\frac{dx^\nu}{ds},
\end{equation}
and expressing: $\partial_\nu T= \frac{\partial T}{\partial \rho}\partial_\nu \rho$ where the trace of the energy-momentum, according to Eq. \eqref{Trho} is given by:
\begin{equation}\label{eq:traza}
	T=4\mathcal{L}_m-3\rho\frac{d \mathcal{L}_m}{d\rho},
\end{equation} 
the geodesic equation of this metric-affine \( f(R,T) \) theory is provided by
\begin{equation}\label{eq:geod}
\frac{d^2x^\mu}{ds^2}+\Gamma^\mu\,_{\lambda\nu}\frac{dx^\lambda}{ds}\frac{dx^\nu}{ds}=f^\mu,
\end{equation}
where the extra force \( f^\mu \) is given by
\begin{eqnarray}\label{eq:force}
	f^\mu&=&-\nabla_{\nu}^{(g)}\ln \left[\frac{d \mathcal{L}_m}{d\rho}+\frac{f_T}{2\kappa^2}\left(\frac{d \mathcal{L}_m}{d\rho} - 3\rho\frac{d^2 \mathcal{L}_m}{d\rho^2}\right)\right] \nonumber \\
	&&
	\times (g^{\mu\nu} +u^\mu u^\nu).
\end{eqnarray}
In other words, in this formulation the particles follow geodesic
trajectories if and only if \( f^\mu = 0 \).

To illustrate the above statement, note that for the case of dust,
this extra force takes the following expression:
\begin{equation} \label{eq:forcep}
f^\mu_{dust}=-(g^{\mu\nu} + u^\mu u^\nu)\nabla_{\nu}^{(g)}\ln \left(1+\frac{f_T}{2\kappa^2}\right) 
\end{equation}
It is clear from this last relation that the extra force vanishes only for
the case $f_T=0$, i.e., \( f(R,T) \) is only a function of \( R \), which
coincides with the standard metric-affine approach of \( f(R) \) gravity
(see for example the direct calculation of this made in \cite{koivisto06}).

\section{Weak field, slow-motion limit} \label{sec:wf}

To investigate the weak field limit of these theories, we start from the conformal relation (\ref{eq:conf}), whose perturbation can be expressed as
\begin{equation}
\delta g_{\mu\nu}=\frac{\delta h_{\mu\nu}}{f_R} - \frac{h_{\mu\nu}}{f_R^2} \delta f_R \,.
\end{equation}
Now let us introduce perturbations upon a Minkowski background, namely, $h_{\mu\nu} \approx \eta_{\mu\nu}+\bar{t}_{\mu\nu}$ and $g_{\mu\nu} \approx \eta_{\mu\nu} + t_{\mu\nu}$, where $\bar{t}_{\mu\nu} \ll \eta_{\mu\nu}$ and $t_{\mu\nu} \ll \eta_{\mu\nu}$. This means that, at the background level via the conformal relation above, one has $f_R \approx 1$ (but $\delta f_R \neq 0$). On the other hand, using the standard gauge choice $\partial_{\lambda} (\bar{t}^{\lambda}_{\mu} -\frac{\bar{t}}{2} \delta_{\mu}^{\lambda})=0$ one finds that $R_{\mu\nu} (\eta_{\mu\nu}+\bar{t}_{\mu\nu}) \approx -\frac{1}{2} \Box \bar{t}_{\mu\nu}$, where $\Box$ is the D'Alambertian (in flat space). After noting that $\delta {R^\mu}_{\nu}(h) \approx \eta^{\mu\alpha}\delta R_{\alpha \nu}$, inserting these results into the field equations (\ref{eq:Riccieqs}) one arrives at
\begin{equation} \label{eq:perteqs}
-\frac{1}{2} \Box \bar{t}_{\mu\nu} = \kappa^2 \left(\tau_{\mu\nu}+\frac{f}{2\kappa^2} \eta_{\mu\nu} \right) \,.
\end{equation}

Limiting ourselves to the nonrelativistic source limit ($P\to 0$), one can compute $\tau_{\mu\nu} \approx \rho \left(1+f_T/\kappa^2 \right)u_{\mu}u_{\nu}$, from where the perturbed field equations (\ref{eq:perteqs}) read
\begin{equation} \label{eq:pertform}
-\frac{1}{2} \vec{\nabla} \bar{t}_{\mu\nu} \approx \kappa^2 \rho \left(1+\frac{f_T}{\kappa^2} \right)u_{\mu}u_{\nu} + \frac{f}{2} \eta_{\mu\nu} \ .
\end{equation}
Given that the background solution is flat Minkowski space and that $\rho$ represents the leading order contribution from the matter sector, the term proportional to $f_T$ in the above expression must be regarded as higher order and, thus, negligible to this order of approximation. Nonetheless, we will keep track of this contribution in the equations by defining the quantity
\begin{equation}
\rho_T= \rho \left(1+\frac{f_T}{\kappa^2} \right) \,.
\end{equation}
Assuming a standard structure for the metric perturbations
\begin{equation}\label{eq:permatrix}
\bar{t}_{\mu\nu}=\begin{pmatrix}
 -2\bar{\phi}_N & \hat{0}_{3 \times 1}    \\
\hat{0}_{1 \times 3} & \bar{\psi} \delta_{ij}\hat{I}_{3 \times 3} \\
\end{pmatrix} \ ,
\end{equation}
where $\hat{I}$ and $\hat{0}$ are the identity and zero matrices, respectively, then the $(0,0)$ component of the perturbation equations (\ref{eq:pertform}) reads
\begin{equation} \label{eq:per00}
\vec{\nabla}^2 \bar{\phi}_N \approx \kappa^2 \rho_T - \frac{f}{2} \ .
\end{equation}
Now, given that $\delta g_{\mu\nu}=\bar{t}_{\mu\nu}-\eta_{\mu\nu}\delta f_R$  and $\delta f_R=f_{RR} \delta R$, one can write the Newtonian potential $\phi_N\equiv -\delta g_{00}/2$ using Eq. (\ref{eq:traceEq}) as
\begin{equation}
\bar{\phi}_N=\phi_N+\lambda \rho \ ,
\end{equation}
where $\lambda\equiv (f_R-R f_{RR})^{-1}f_{RR}\kappa^2/2$ is evaluated in vacuum. This leads to the following modified Poisson equation for metric-affine $f(R,T)$ theories:
\begin{equation} \label{eq:Poisson}
\vec{\nabla}^2 {\phi}_N \approx \kappa^2 \rho_T - \frac{f}{2} -\lambda\vec{\nabla}^2 \rho \ .
\end{equation}
Given that in this equation $f(R,T)$ is a function of $\rho$ and $P$, using the notation $\kappa^2\tilde{\rho}/2\equiv \kappa^2 \rho_T - f/2$, this expression boils down to the usual result in the GR limit,  which allows to write
\begin{equation}
\phi_N=\frac{\kappa^2}{8\pi}\int d^3\vec{x}' \frac{\tilde{\rho}(t,\vec{x}')}{|\vec{x}-\vec{x}'|}-\lambda \rho \ .
\end{equation}
This modified Newtonian potential is formally identical to that found in the weak field limit of the Eddington-inspired Born-Infeld (EiBI) theory of gravity (see the recent review \cite{blor17a}, Sec.~3) and, therefore, the implications derived from it might be similar except, perhaps, due to new effects arising from the redefinitions introduced above. These similarities are expected, in particular, in nonrelativistic stellar models.

\section{Some applications} \label{sec:mod}

\subsection{Stellar structure equations}

The weak field equations derived above were useful to establish some relations between the physics of metric-affine $f(R,T)$ models and other gravity theories such as the EiBI model. In this section we derive the complete Tolman-Oppenheimer-Volkov (TOV) equations for hydrostatic equilibrium to show that the metric-affine version of $f(R,T)$ theories studied in this work does introduce different physics in the full relativistic regime. For this purpose, we consider the nonconservation equation (\ref{eq:contau}) applied to a perfect fluid (\ref{eq:pftumunu}) to find
\begin{equation}
\partial_r P=-\frac{\left(1+\kappa^{-2}f_T\right)(\rho+P)}{\left[1+\frac{2}{\kappa^2}\left(f_T+P\partial_P f_T+\frac{1}{4}f_T\partial_P T\right)\right]  } u^\alpha \nabla_\alpha u_r \ .
\end{equation}
 In the $f_T\to 0$ limit, this equation recovers the usual structure equation of GR and of metric theories of gravity with no matter-curvature couplings. For static, spherically symmetric configurations, only the radial derivative equation survives and one finds that $u^\alpha \nabla_\alpha u_r = \Gamma^t_{tr}=A_r/2A$, where $g_{tt}=-A(r)$. The resulting TOV equation thus takes the form
\begin{equation}
\partial_r P=-\frac{\left(1+\kappa^{-2}f_T\right)(\rho+P)}{ \left[1+\frac{2}{\kappa^2}\left(f_T+P\partial_P f_T+\frac{1}{4}f_T\partial_P T\right)\right]  } \frac{A_r}{2A} \ .
\end{equation}
The weak field limit obtained  in the general case above follows from this equation by taking
\begin{equation}
(\rho+P)\approx \rho \ , \qquad \kappa^{-2}f_T\to 0 \ ,
\end{equation}
and
\begin{equation}
A_r\approx 2\left[\frac{\kappa^2 M(r)}{8\pi r^2}-\lambda \rho_r\right] \,,
\end{equation}
with $M(r)=\int^r d^3\vec{x} x^2 \tilde{\rho}(t,\vec{x})$. After setting specific $f(R,T)$ models these equations allow to solve any scenario of interest in this context.

\subsection{Electromagnetic fields}

Let us consider now the case of an electromagnetic field. For a Maxwell field, described by the Lagrangian density $\mathcal{L}_m=-\frac{1}{16\pi} F_{\mu\nu}F^{\mu\nu}$, where $F_{\mu\nu}=\partial_{\mu}A_{\nu}-\partial_{\nu}A_{\mu}$ is the field strength tensor, from Eq. (\ref{eq:Thetaexp}) one finds that
\begin{equation}
\Theta_{\mu\nu}=-T_{\mu\nu}=-\frac{1}{4\pi}\left(F_{\mu\alpha} {F_\nu}^{\alpha} - \frac{1}{4}g_{\mu\nu}F_{\alpha\beta}F^{\alpha\beta}\right) \ .
\end{equation}
From Eq. (\ref{eq:tau}) this result yields the cancellation of the $f_T$ contributions which, together with the tracelessness of Maxwell's energy-momentum tensor, implies that any solutions for these matter fields will coincide with those of GR regardless of the $f(R,T)$ theory chosen.

In order to find nontrivial new physics associated with electromagnetic fields,  one must go beyond Maxwell's theory and consider instead nonlinear electrodynamics theories.  In this case, defining the matter sector as
\begin{equation}
\mathcal{S}_m=\frac{1}{8\pi} \int d^4x \sqrt{-g} \, \varphi(X) \ ,
\end{equation}
where $\varphi(X)$ is a function of the field invariant $X=-\frac{1}{2}F_{\mu\nu}F^{\mu\nu}$ specifying the model of nonlinear electrodynamics\footnote{Functions of a second field invariant, $Y=-\frac{1}{2}F_{\mu\nu}F^{\star \mu\nu}$, built out of the dual field strength tensor, $F^{* \mu \nu} \equiv \frac{1}{2} \varepsilon^{\mu\nu\alpha\beta}F_{\alpha\beta}$, are also possible, but for simplicity we shall not consider them here.} (Maxwell electrodynamics corresponding to $\varphi(X)=X$). The corresponding energy-momentum tensor reads
\begin{equation}
T_{\mu\nu}=\frac{1}{4\pi}\left(\varphi_X F_{\mu\alpha}{F_\nu}^{\alpha} +\frac{\varphi}{2}g_{\mu\nu} \right) \ ,
\end{equation}
where $\varphi_X \equiv \partial \varphi / \partial X$. In this case it is easy to find that
\begin{equation}
\Theta_{\mu\nu}=-T_{\mu\nu}+\frac{1}{2\pi} X\varphi_{XX}F_{\mu\alpha}{F_\nu}^{\alpha}\,,
\end{equation}
and
\begin{equation}
\tau_{\mu\nu}=T_{\mu\nu}-\frac{f_T}{2\pi \kappa^2} X \varphi_{XX}F_{\mu\alpha}{F_\nu}^{\alpha}\,.
\end{equation}
The new $f_T$ contributions induce modifications as compared to GR solutions, as we shall see at once with an explicit example.

Let us focus on (electro-)static, spherically symmetric solutions, for which the only nonvanishing component of the field strength tensor is $F_{tr} \neq 0$. In this case, the matter energy-momentum tensor reads
\begin{eqnarray} \label{eq:Amatrix}
{T_\mu}^{\nu}&=& \frac{1}{4\pi}
\left(
\begin{array}{cc}
\left[-X\varphi_X + \frac{\varphi}{2} \right] \hat{I} &  \hat{0} \\
\hat{0} & \frac{\varphi}{2}\hat{I}   \\
\end{array}
\right),
\end{eqnarray}
where now $X=-F_{tr}F^{tr}$, while the conserved energy-momentum tensor takes the form
\begin{eqnarray}
 {\tau_\mu}^{\nu}&=& \frac{1}{4\pi}
\left(
\begin{array}{cc}
\left[-X\varphi_X + \frac{\varphi}{2} +2\frac{f_T}{\kappa^2}X^2 \varphi_{XX} \right] \hat{I} &  \hat{0} \\
\hat{0} & \frac{\varphi}{2}\hat{I}   \\
\end{array}
\right) ,  \label{eq:em}
\end{eqnarray}
where $\hat{I}$ and $\hat{0}$ are the $2 \times 2$ identity and zero matrices, respectively. To proceed further and find solutions we need to specify an $f(R,T)$ model. For simplicity and to illustrate the general procedure to solve the field equations, let us choose the simple model $f(R)=R+\epsilon T$, where $\epsilon$ is some parameter\footnote{Cosmological FRW-type solutions of this model can also be easily worked out, with the result that, for dust, the scale parameter behaves as $a(t) \propto t^{\alpha}$, where $\alpha=\frac{2}{3}\left(\frac{\kappa^2+3\epsilon/2}{\kappa^2+\epsilon}\right)$. This is a similar result as that obtained in the metric formulation of these theories \cite{Harko:2011kv}.}. From the trace equation (\ref{eq:traceEq}) one finds that $R=-(\kappa^2 + 2\epsilon)T - \frac{f_T}{\pi \kappa^2} X^2 \varphi_{XX}$ and inserting this result into the field equations (\ref{eq:Riccieqs}), a bit of algebra yields
\begin{eqnarray}
{R^\mu}_{\nu}(h)=\frac{\kappa^2}{f_R^2}  \left(
\begin{array}{cc}
\bar{\varphi} \, \hat{I} &  \hat{0} \\
\hat{0} & \left(\bar{\varphi}+\bar{\varphi}_X\right) \hat{I}   \\
\end{array}
\right) \,,  \label{eq:feem}
\end{eqnarray}
where we have defined the quantities
\begin{eqnarray}
\bar{\varphi}&=&-\frac{1}{4\pi} \left(\frac{\varphi}{2} + \frac{\epsilon}{\kappa^2}(\varphi-X\varphi_X) \right) \,, \\
\bar{\varphi}_X &=& \frac{1}{4\pi}  \left( X\varphi_X - \frac{\epsilon}{\kappa^2}\, 2X^2 \varphi_{XX} \right) \,,
\end{eqnarray}
for notational convenience.

To solve this kind of field equations in metric-affine gravities one usually introduces two different line elements, one for $g_{\mu\nu}$ and another one for $h_{\mu\nu}$, and then makes use of the conformal transformation (\ref{eq:conf}) to work out the relations among the functions on each line element. However, for the model chosen here, $f_R=1$, and such line elements become the same. Let us thus propose an ansatz for a static, spherically symmetric line element of the form
\begin{equation}
ds^2=-A(r)e^{2\psi(r)}dt^2-\frac{1}{A(r)}dr^2+r^2d\Omega^2 \ ,
\end{equation}
where $\{A(r), \psi(r)\}$ are functions of the radial coordinate $r$ and $d\Omega^2=d\theta^2+ \sin^2 \theta d\phi^2$ is the angular element on the unit $2$-spheres. From the combination ${R^t}_t-{R^r}_r=0$ of the field equations (\ref{eq:feem}) it follows that $\psi(r)={\rm constant}$, which can be set to zero without loss of generality. As for the component ${R^\theta}_{\theta}=\frac{1}{r^2} \left[1-A(r)-rA_r\right]$ on the left-hand side of Eqs. (\ref{eq:feem}), introducing a standard mass ansatz of the form $A(r)=1-2M(r)/r$, it can be solved as (recall that $X=X(r)$)
\begin{equation} \label{eq:massol}
M(r;\epsilon)=M_0-\frac{\kappa^2}{2}\int_r^{\infty} dR R^2 \left[\bar{\varphi}(X)+\bar{\varphi}_X(X) \right] \ ,
\end{equation}
where $M_0$ is an integration constant identified as Schwarzschild's mass. The next step to provide explicit solutions would be to supply a specific function $\varphi(X)$, i.e., to choose any of the nonlinear models of electrodynamics studied in the literature, for instance, in the context of spherically symmetric solutions in GR, see e.g. \cite{NED-GR,NED-GR2,NED-GR3,NED-GR4,NED-GR5,NED-GR6,NED-GR7,NED-GR8}. Once given, the resolution of the corresponding matter field equations, $\nabla_{\mu}(\varphi_X F^{\mu\nu})=0$, would provide the explicit expression of $X(r)$ needed to carry out the integral in Eq. (\ref{eq:massol}), thus closing the problem. The analysis of this kind of models and solutions could open new avenues in the investigation of outstanding problems in this context, such as the singularity avoidance within nonlinear electrodynamics coupled to gravity, paralleling previous analysis carried out in the context of metric-affine $f(R)$ theories, see e.g. \cite{Bambi:2015zch}.

\subsection{Scalar fields}

Scalar fields represent yet another example suitable for investigation within these theories and yielding nontrivial new dynamics. Defining in this case the Lagrangian density as $\mathcal{L}_m=\frac{1}{2} (g^{\mu\nu}\partial_{\mu}\phi\partial_{\nu}\phi + 2V(\phi))$ where $V(\phi)$ is the potential, one finds $\Theta_{\mu\nu}=-2T_{\mu\nu}+g_{\mu\nu}\mathcal{L}_m$, and the effective energy-momentum tensor reads
\begin{equation}
{\tau^\mu}_{\nu}={T^\mu}_{\nu}\left(1+\frac{f_T}{\kappa^2}\right)-\frac{f_T}{\kappa^2}\mathcal{L}_m {\delta^\mu}_{\nu} \,.
\end{equation}
Likewise the electromagnetic field case above, setting specific $f(R,T)$ models and working out the corresponding field equations one may find $f_T$-corrections to GR solutions, which brings about new possibilities. For instance, free ($V=0$) geonic solutions of the kind found in Ref. \cite{AOR17} in the context of Eddington-inspired Born-Infeld gravity should also be possible in metric-affine $f(R,T)$ theories.

\section{Conclusion} \label{sec:con}

In this work we have derived the field equations of $f(R,T)$ theories with independent metric and affine connection (metric-affine approach). We have found that for matter sources not coupled to the connection (for which the torsion degrees of freedom are trivial \cite{Afonso:2017bxr}), the symmetric part of the connection can be written as the Levi-Civita connection of an auxiliary metric conformally related to $g_{\mu\nu}$ via the matter sources, and that the resulting field equations can be formally written in the same way as those of metric-affine $f(R)$ theories once an effective energy-momentum tensor is defined. These equations impose the nonconservation of the energy-momentum tensor, therefore entailing nongeodesic motion and the appearance of a fifth force, which has a nontrivial impact for the physics of compact objects and relativistic stars. For nonrelativistic stellar objects, the dynamics is qualitatively similar to that found in the EiBI model, for which there exists extensive literature \cite{blor17a}.

After having under control the basic framework for metric-affine $f(R,T)$ gravity, we have introduced the main elements for some applications. When coupled to perfect fluids, the nonconservation equation introduces novelties in the hydrodynamical equilibrium equation in the full nonrelativistic regime, with expected non-negligible consequences for compact objects in this context. When coupled to electromagnetic fields, we have shown that these theories yield the same solutions as GR unless a nonlinear theory of electrodynamics is considered, where the problem of non-singular black holes can be tackled from a different perspective, and similar comments apply to scalar fields.

In summary, the primer $f(R,T)$ gravity in the metric-affine formalism developed in this work opens new avenues of research and the possibilities to explore new physics in this context are huge. Further research is expected in these and other directions in the future, on which we hope to report soon.

\section*{Acknowledgments}
We thank Jimmy Wu,  Guangjie Li and Tiberiu Harko for calling our attention to an
error in equation~(29) in the first version of our manuscript.
E.B. and S.M. acknowledge economic support from CONACyT (517586, 26344) and further support by DGAPA-UNAM and CONACyT grants (PAPIIT IN112616 and CB-2014-1~\#240512). F.S.N.L. is funded by the Funda\c{c}\~ao para a Ci\^encia e a Tecnologia (FCT, Portugal) through an Investigador FCT Research contract No.~IF/00859/2012. G.J.O. is funded by the Ramon y Cajal contract RYC-2013-13019 (Spain). DRG is funded by the FCT postdoctoral fellowship No.~SFRH/BPD/102958/2014. F.S.N.L. and D.R.G. also acknowledge funding from the research grants UID/FIS/04434/2013 and No.~PEst-OE/FIS/UI2751/2014. This work is supported by the Spanish project FIS2014-57387-C3-1-P (MINECO/FEDER, EU), the project H2020-MSCA-RISE-2017 Grant FunFiCO-777740, the project SEJI/2017/042 (Generalitat Valenciana), the Consolider Program CPANPHY-1205388, and the Severo Ochoa grant SEV-2014-0398 (Spain). This article is based upon work from COST Action CA15117, supported by COST (European Cooperation in Science and Technology).




\end{document}